# Threats and Countermeasures of Cyber Security in Direct and Remote Vehicle Communication Systems

**Subrato Bharati[1], Prajoy Podder[2], M. Rubaiyat Hossain Mondal[3], Md. Robiul Alam Robel[4]**

[1,2,3] Institute of Information and Communication Technology,
Bangladesh University of Engineering and Technology
Dhaka, Bangladesh
[1]*subratobharati1@gmail.com*, [2]*prajoypodder@gmail.com*, [3]*rubaiyat97@yahoo.com*

[4] Department of Computer Science and Engineering,
Cumilla University, Cumilla, Bangladesh.
*alam.robel@gmail.com*

*Abstract*: Traffic management, road safety, and environmental impact are important issues in the modern world. These challenges are addressed by the application of sensing, control and communication methods of intelligent transportation systems (ITS). A part of ITS is a vehicular ad-hoc network (VANET) which means a wireless network of vehicles. However, communication among vehicles in a VANET exposes several security threats which need to be studied and addressed. In this review, firstly, the basic flow of VANET is illustrated focusing on its communication methods, architecture, characteristics, standards, and security facilities. Next, the attacks and threats for VANET are discussed. Moreover, the authentication systems are described by which vehicular networks can be protected from fake messages and malicious nodes. Security threats and counter measures are discussed for different remote vehicle communication methods namely, remote keyless entry system, dedicated short range communication, cellular scheme, Zigbee, Bluetooth, radio frequency identification, WiFi, WiMAX, and different direct vehicle communication methods namely on-board diagnosis and universal serial bus.

## I. Introduction

Vehicular networking for intelligent transportation system (ITS) can be considered as an emerging technology. It has attracted the attention of industrialist, academia in USA, Japan, and Europe. The association of electric technology for automobile traffic and driving (Japan) popularly known as JSK established the idea of vehicular communication and networking in the early 1980s. The formal development process began in 1990s in USA. The vehicles used in modern times can no longer be appraised as just mechanical arrangements, including 100 million lines of code in the total architecture, more sophisticated than a modern operating system. One part of ITS is the vehicular ad-hoc network (VANET) which is a wireless network of vehicles. Within a VANET, a vehicle can communicate with other vehicles and with the traffic light or other road side infrastructure. These vehicles exchange important information for example, traffic condition, accident warning, curvature of the roads, etc.

In a VANET, the main role is performed by vehicles which have computation facilities and can even connect to remote devices like smart phones. A smart vehicle can now provide different facilities such as safety information, weather and navigation information to the driver. A vehicle can also communicate with other nearby vehicles, traffic lights or other road side infrastructures. These facilities have brought a number of advantages to the occupants, but there is also the introduction of hackers to track the vehicle and create different forms of risks.

The severity of cyber threats in vehicular communication system can be categorized into three ways:

(i) High threats: On-board diagnostics (OBD) hack, V2V hack, V2I hack, GPS spoofing, MITM.

(ii) Medium threats: key fob hacking, attack on control area network (CAN) bus, entertainment system hacking.

(iii) Low threats: Airbags hacking, brakes hacking

Figure 1 illustrates the common requirements needed to establish a highly secured vehicular communication system.

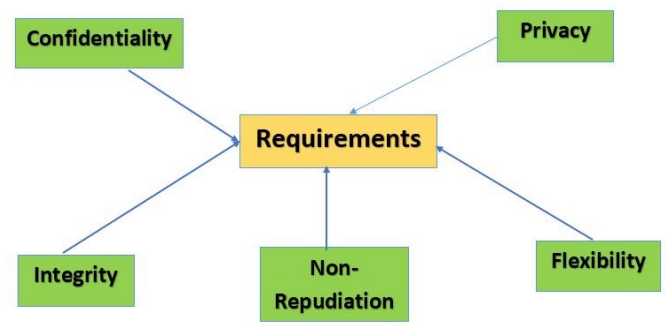

**Figure 1.** Security Requirements

Threats occurred in vehicular communications can be discussed by the autonomous vehicular sensing-communication-control (AutoVSCC) model. In sensing layer, eavesdropping or spoofing may occur on vehicle sensors. In communication layer, inter-vehicular and intra-vehicular communication are included. In communication layer, eavesdropping may occur. Messages can be manipulated between cars and roadside infrastructure. Threats to the sensing layer and communication layer can affect the control layer. Speed of the cars and steering control can be hampered due to the threats of the control layer [1-3]. Despite a number of research works, there are still many security challenges in vehicle communication systems. This paper provides a comprehensive review of the security threats and countermeasures for direct and remote vehicle communication. The rest of the paper is organized as follows. Section II discusses different remote vehicle communication methods namely, remote keyless entry system, dedicated short range communication, cellular scheme, Zigbee, bluetooth, RFID, WiFi, and WiMAX. The security threats and countermeasures of these systems are also described in Section II. Section III presents the security issues and their remedies for direct vehicle communication methods namely OBD and USB. Finally, the concluding remarks are provided in Section IV.

## II. Remote Communication Technologies

Remote communication systems can be used within vehicle-to-everything communication. With this communication, messages can be passed between onboard units and roadside units. Weak Points can be focused by attackers in the area of remote communication technologies for the purpose of tampering wickedly with a car's or bus's functioning from a suitable distance. Therefore, attackers need not connect external devices to vehicle ports for achieving entry to vehicle communication system architectures [4-6]. There are some limitations in remote communication technologies and attackers can detect the limitations. Figure 2 classifies the remote communication technology into several categories such as keyless entry system (RKE), wireless access in vehicular system, Zigbee, radio frequency identification (RFID), WiMAX, Wi-Fi, etc.

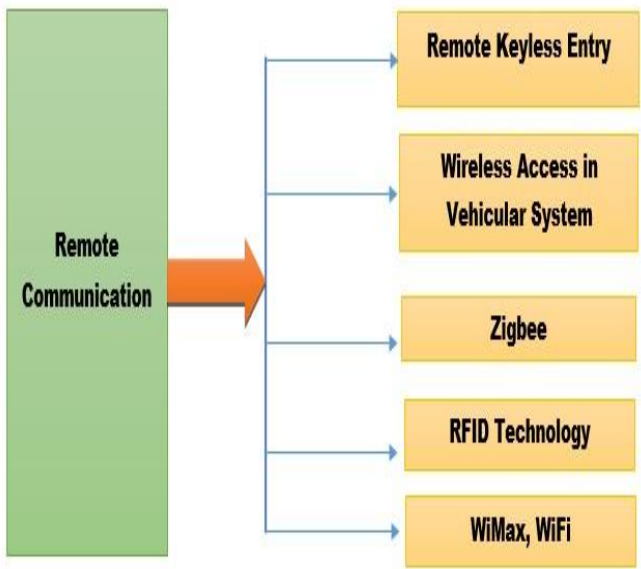

**Figure 2.** Types of Remote Communication

### A. Remote Keyless Entry System

Remote keyless entry system can also be called remote central locking. This system refers to an electronic remote controlled lock. This lock can be activated not only by a hand operated device but also by proximity. The lock is controlled by a keypad. The keypad is located at or near the driver's door. Ford and Lincoln car brand use this system.

#### 1) Threats

There are a number of vehicle manufacturers that use cryptographic RKE keys. One example is the VW Group using rolling code schemes [7]. However, the rolling codes are vulnerable to eavesdrop attacks and RKE cloning.

At the 2015 DEFCON hacker convention, it was shown that attackers could jam the signal to the RKE receiver and use a different receiver to store RKE codes [7]. Tewari et al. [8] provided an extensive listing of attacks to which RKE systems are vulnerable and notes that RKE systems' three main vulnerabilities are their frequent use of outdated devices and techniques, weak cryptographic schemes, and implementation faults. Glocker et al. [9] also discussed the potential attacks on RKE systems, and Liu et al. [10] examined the vulnerabilities of and potential attacks on the Hitag2 cipher, which is used in many RKE systems.

#### 2) Countermeasures

S. Van De Beek et al. [11] designed a robust receiver. According to them, the receiver is less susceptible to jamming attacks. Hamadaqa et al. [12] described a Secret Unknown Cipher (SUC) methods that would allow RKE controllers to be cloneresistant. Garcia and Oswald [7] presented vulnerabilities in keyless entry methods. They illustrated Hitag2 rolling code structure. Hitag2 can be used in vehicles made by Alfa Romeo, Ford, Chevrolet, Peugeot, etc. Zhang et al. [13] presented effective k-Means Authentication 2 (EKA2). EKA2 validates the revocation status of digital certificates with the help of clustering approach for authenticating RKE devices in order to communicate with the vehicle. Lee et al. [14] provided a Rhythm Key-based approach to encrypting RKE communications.

### B. DSRC

Dedicated Short Range Communication is a type of wireless communication technology. The technology is 802.11p-based. This communication provides robust security, direct communication with high amount of speed between vehicles such as buses, cars and the adjacent infrastructure. This system does not involve any cellular infrastructure. IEEE 802.11p standard states enrichments of ITS related applications. Figure 3 shows a basic diagram of DSRC used in vehicular communication. In this figure, RSU denotes the road side unit and V2I means vehicle to infrastructure communication system.

The operating frequency band of DSRC is 5.9 GHz. The FCC dedicated a bandwidth of 75 MHz for DSRC technology in the 5.9 GHz band in 2004. WAVE (Wireless Access in Vehicular Environments) systems include OBUs and RSUs [15]. Y. Li [15] illustrated a summary of DSRC and WAVE.

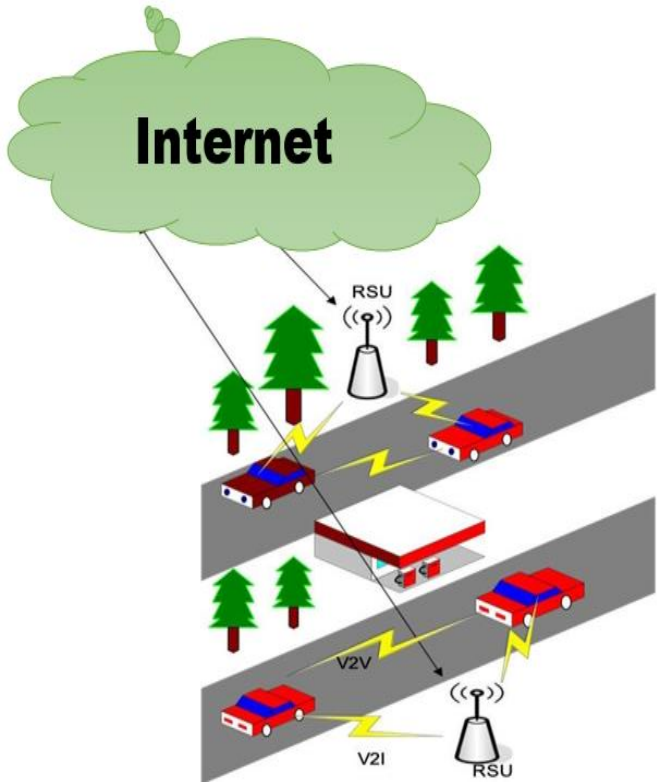

**Figure 3.** DSRC System (Adapted from [73])

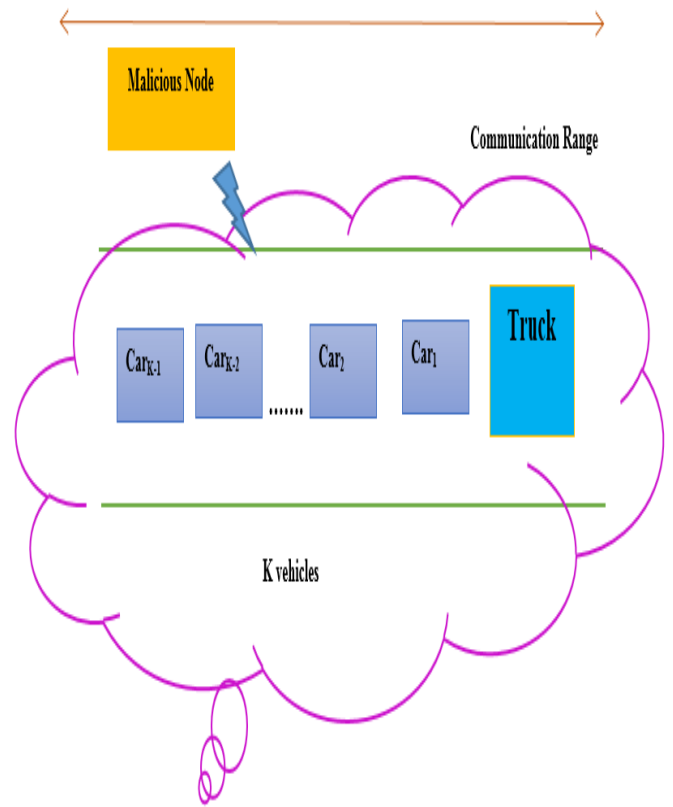

**Figure 4.** Platooning Scenario

*1) Threats*

Laurendeau et. al. [16] identified some important threats such as malware, DoS, spamming, black hole, location tracking GPS spoofing, etc. These threats are barrier to the security of DSRC or WAVE based networks. They used ETSI's threat methods for threat identification. They discussed probable solutions to the critical threats. According to them, DoS attack can be considered as major threats. S. C. Urgen [17, 18] proposed a simulation platform which supports both DSRC and Visible light communication in platoon. Urgen also analyzed the stability of platoon under false packet condition and duplication of security. Visible light Communication (VLC) a form of optical wireless communication [19] is susceptible to attacks owing to the use of antennas that are omnidirectional. The authors expanded that work in [20] and pushed for a fusion of VCL and DSRC protocol in order to mitigate jamming, packet injection or other attacks in platoon. W. Whyte [21] identifies the threats encountered by WAVE service system. They observed that wave service system is susceptible to not only availability threats but also privacy threats. Privacy threats occur from Eavesdropping attacks.

*2) Countermeasures*

N. Lyamin proposed a real-time detector that can clog Distributed of service (DoS) attacks in the VANET platoons [22]. Platooning scenario is shown in figure 4. The detector had been validated in terms of two condition within a time range. They are (i) detection and (ii) false alarm probabilities.

Huong Nguyen-Minh et al. [23] proposed a detection method of distinguishing packet loss due to several reasons such as normal collisions and jamming attacks. They also described observation and estimation model. They described the impact of wrong estimation at the time of probability detection graphically. Their simulation results showed that when attack probability is high then detection probability is high and collisions condition is low.

Nidhi Gambhir et. al. [24] proposed a novel network model with the generation of RSA public or private key pair in order to initiate connection in the VANET. Three common encryption techniques such as DES (Data encryption standard), AES (Advanced encryption standard) and RSA algorithm were discussed and their performance was compared with the proposed method in terms of computation time of encryption decryption process, buffer size in bytes, encryption and decryption throughput in Byte/second. Communication over IEEE 802.11p standard can be accomplished by their proposed hybrid technique.

*C. Cellular Technology*

DSRC has some limitations. The ability of DSRC to support reliable and efficient vehicle-to-vehicle communications is poor especially in high vehicle density situations. Additionally, the allocated DSRC radio spectrum is not estimated to meet the high data traffic demand for in-vehicle internet access. DSRC has short range. The network path in which the packets route to or from the gateway reduce the appropriateness of DSRC for any Vehicle-to-everything communication. For this case, DSRC requires low latency data dissemination on a large road segment. There is also limitation of DSRC based CSMA/CA technique, which is the main contention-based MAC scheme employed by DSRC standards, such as IEEE 802.11. The currently used cellular technology in 4th generation (4G) is the Long-Term Evolution (LTE) and LTE-advanced. These systems enable high speed wireless cellular communication for a large number of mobile users with the use of orthogonal frequency division multiplexing (OFDM) [5, 19, 25-28], and its multiple access version known as OFDMA. The next generation of cellular systems 5G and 6G are expected to use derivatives of OFDM [27] in order to enable even higher data rates for more users than 4G.

Pure-DSRC vehicle to everything (V2X) communication is gaining popularity with the amalgamation of cellular technologies for various applications. Figure 5 shows a scenario of V2X communication system. Several reasons exist for this interest:

a) High network capacity with high bandwidth

b) Wide cellular coverage range,

c) Mature technology, which eases the implementation and accelerates the deployment of V2X communications.

In spite of having these benefits, there are also many barriers. The barriers can limit the ability of cellular technology to support reliable V2X communications.

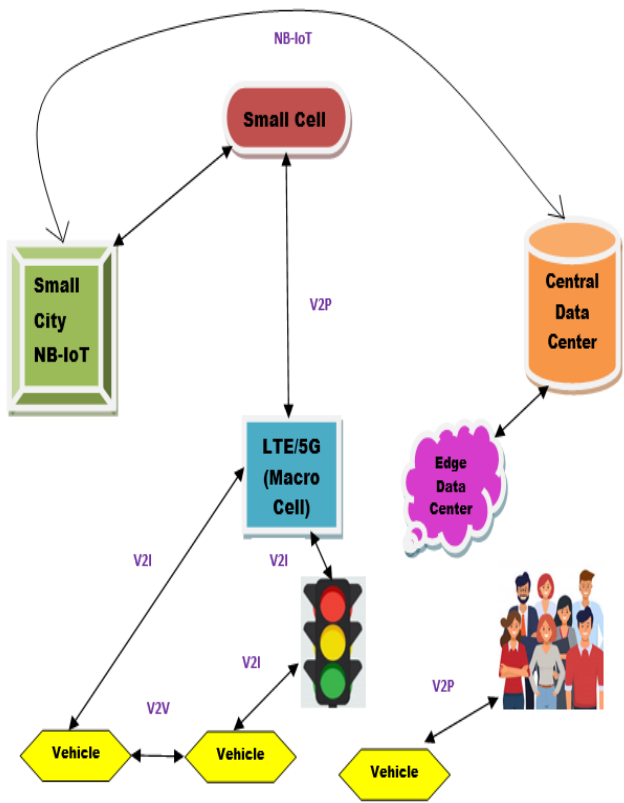

**Figure 5.** V2X communications

Cellular V2X (C-V2X) has been developed by 3GPP1. This technology has two different modes [29]. One is the device to device mode which includes vehicle-to-vehicle (V2V), vehicle-to-pedestrian (V2P) and vehicle-to-roadway infrastructure (V2I). In this mode, there is direct communication and thus scheduling is not required for networking. The second mode is the device-to network or vehicle-to-network (V2N) mode. This mode enables a number of cloud services with the use of cellular communication.

*1) Threats*

Jin Cao et al. [30] made a number of contributions to the security characteristics of the LTE and LTE-Advanced networks. They presented a summary of the security functionality of the LTE and LTE-Advanced networks with the exploration of the security weaknesses existing in the architecture of the LTE and LTE-Advanced networks.

There are some security risks in LTE networks such as the vulnerability to the injection, modification, eavesdropping attacks [31]. So, it has more privacy risks than GSM and UMTS [32].

Ravishankar Borgaonkar proposed a new privacy attack against the variants of the authentication and key management (AKA) protocol. Their proposed attack breaches the privacy of the subscribers' more rigorously than the common attacks do. They show that moderately learning SQN leads to activity monitoring attacks [33]. LTE network suffers from various attacks such as spoofing the IP address, DoS attacks, worms, viruses, etc. [32]. Since LTE femto cell is launched, HeNBs (Home enodeB) are attained by the attacker. The attackers can modify the functionality of the cell according to his own requirements [34].

David Rupprecht et al. [35] focused on two types of security characteristics:

(a) Encrypting user data.

(b) Authentication of the network.

A framework was developed in [35] for analyzing different LTE devices. They show some important characteristics after analyzing the device. They are:

  i) An LTE network can impose on using no encryption

  ii) None of the tested devices informs the user about unencrypted message.

  iii) MITM attack against an LTE device that does not fulfill the network authentication requirements.

*2) Countermeasures*

Liyange et al. [36] suggested LTE security features. According to the study [36], software defined networking and network function virtualization concepts can improve the security of LTE. J. Cichonski et al. [37] provided some solutions in order to mitigate the LTE security threats. The study illustrated the use of ciphering indicator that is defined in 3GPP TS 22.101. This indicator informs the status of the user plane confidentiality protection. They also discussed the planning of configuration management, patch management, malware and intrusion detection and implementation throughout the mobile network operator's infrastructure for the purpose of alleviating malware attacks impacting radio access network and core infrastructure, unauthorized operation and access management networks.

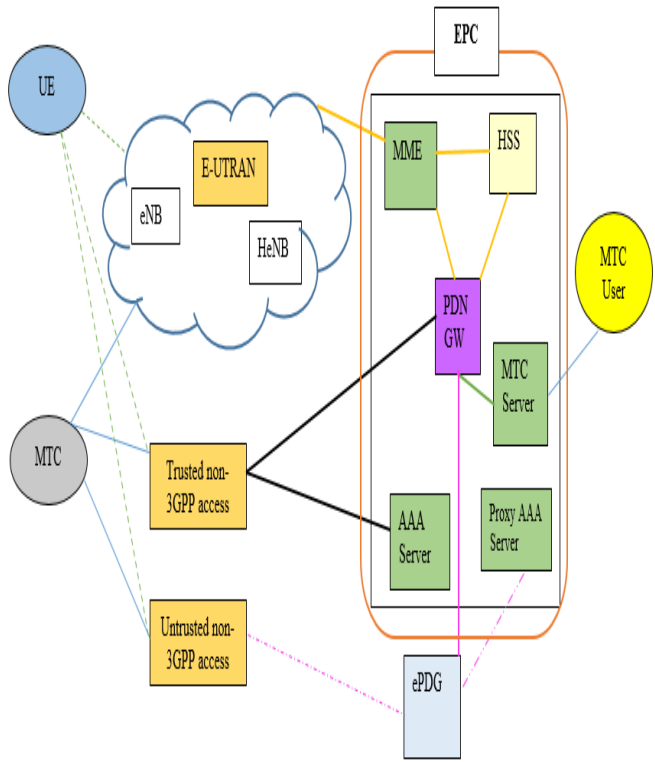

**Figure 6.** Network Architecture of LTE

### D. Zigbee

Zigbee is a wireless standards-based technology for WSN applications in order to address the needs of low-cost, long battery life. Zigbee operates on the IEEE 802.15.4 physical radio specification. It has three operating frequency bands such as 868 MHz, 900 MHz and 2.4 GHz. ZigBee allows mesh networking facility. With this wireless mesh networking facility, ZigBee is applied by routers and receivers in multiple application scenarios [38]. The spread spectrum technique used in Zigbee is direct-sequence spread spectrum. Modulation technique adopted in Zigbee is offset-QPSK. Channel width is 2 MHz and channel spacing is 5 MHz. Yu Lei and Jian Wu [39] applied Zigbee technology into forward collision waring system (FCWS). FCWS uses lasers, sensors, cameras for scanning the road ahead and alerting the driver if the distance to a vehicle ahead is closing too quickly. Sourabh Pawade et al. [40] proposed a Zigbee based advanced driver assistance systems. This system has two units: (i) mobile unit where Zigbee technology was installed (ii) static unit on the road. Mobile unit helps to alert the driver along with audio warnings. Sourabh Pawade et al. also designed a drink and drive prevention system for continuous monitoring the alcohol level of the driver in order to reduce the accidents [40].

*1) Threats*

There are mainly four concepts of Zigbee based security [41].

   a) High level of security

   b) Trust center

   c) AES and DES

   d) Message Integrity code

N. Vidgren et. Al [41] proposed two types of attacks that can harm the ZigBee security. They are:

(i) Sabotage attack: It is based on sabotaging the zigbee end devices when it sends a special signal. The signal makes the devices wake-up constantly until the battery runs out.

(ii) Network Key Sniffing attack: This attack occurs when key exchange system is utilized in ZigBee at the time of using standard security level. The security level can be defined by the ZigBee specification for the purpose of interrupting the network key.

Olayemi Olawumi [42] proposed three practical attacks against the ZigBee security.

(i) Network discovery and device identification attack

(ii) Interception of packets attack

(iii) Replay attack.

Figure 7 illustrates the process of Replay attack.

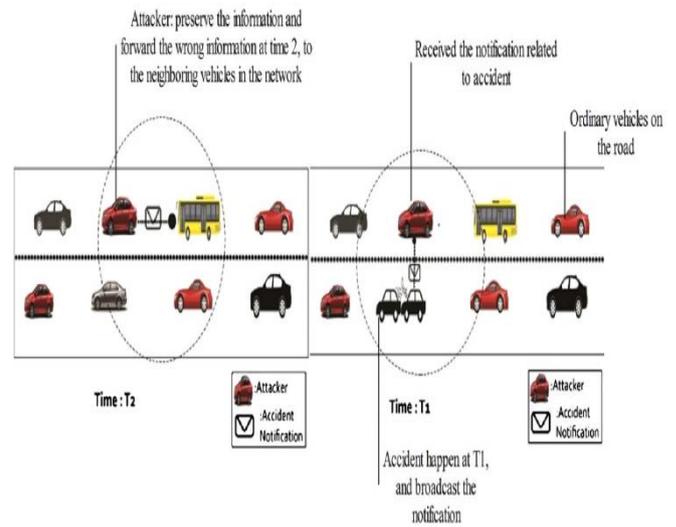

**Figure 7.** Process of Replay attack (Adapted from [74])

*2) Countermeasures*

Cache et al. [43] presented that the best countermeasure against the attack is to recognize the effect of this attack. The attack should be evaluated in own networks for recognizing the information which is visible to the attacker.

Olawumi et al. [42] suggested the use of intrusion detection, the installation of network keys before deployment, and the incorporation of time-stamping within ZigBee's encryption methods to combat replay attacks.

(i) Against interception of packets attack: Standard security level was proposed in [42]. As a result, administrator will get highest level of security.

(ii) Against Replay attack: Fusion of time stamping mechanism with the AES was proposed in order to mitigate the replay attack [42].

### E. Bluetooth

Bluetooth is a type wireless system. This technology is used for exchanging files and messages between fixed and mobile devices. The range of communication of this device is short. Bluetooth uses short-wavelength UHF radio waves. The frequency range is 2.40-2.48 GHz. [44]. Some best cars that have good Bluetooth facility are as follows, Chevrolet Corvette, Mercedes-Benz E-Class, Ram 1500, Toyota Camry, etc. Bluetooth Version 5.2 was released on January 2020. Low energy audio, low energy isochronous channel, low energy power control are the salient features of version 5.2.

*1) Threats*

Several research paper focused on Bluetooth vulnerabilities. According to H. Onishi [45], there are some necessary vulnerabilities faced in the Bluetooth:

(i) Security technique is limited.

(ii) Permission of user is not required at the time of pairing the devices.

(iii) Virus or malware in carry-in devices can destroy the security system.

(iv) After pairing, necessary data such as password, address book can be achieved.

(v) Probability of buffer overflow attack. In buffer overflow, the attacker can overwrite the memory of an application. System can be crashed due to this attack.

(vi) Malicious GNSS signals. GNSS means global navigation satellite system. GNSS has 2 types of vulnerability:

(a) jamming (b) location spoofing

M. Cheah et. al. [46] classified Bluetooth attacks in his paper. Table 1 illustrates various type of Bluetooth attacks.

| Attacks | Threats |
| --- | --- |
| Range Extension | Extension of range by external directional antenna |
| Fuzzing | Inoculation of malicious data |
| Sniffing | Narrow band and wideband receiver is used for dumping raw data |
| DoS | Flooding with many signals that results jamming |
| Malware | Malicious codes or programs injected via interface |
| Unauthorized access of data | Brute forcing PINs, unprotected execution of APIs, utilizing loopholes in the OBEX protocols. |
| Man in the middle attack | Entering anyone in the middle of a V2V communication in order to eavesdrop on or modify data |

*Table 1.* Attacks in Bluetooth

*2) Countermeasures*

H. Onishi offered some technical methods in order to mitigate the Bluetooth vulnerabilities [45].

(i) Penetration testing for mainly V2V communication and interfaces of Bluetooth in order to identify threats.

(ii) Allowance of certified carry-in devices with proper audit in order to connect to vehicles.

(iii) Intrusion detection for detecting abnormal conditions and notifying the driver instantly.

(iv) Availability of ISO-26262 used to define the functional hazards.

*F. Wi-Fi and WiMAX*

Wi-Fi means wireless fidelity. Wi-Fi is a candidate for vehicle to vehicle and vehicle to infrastructure communication. WiMAX is faster than Wi-Fi. WiMAX means Worldwide Interoperability for Microwave Access (WiMAX). WiMAX refers to IEEE 802.16 standard. WiMAX has several features such as low-latency, Quality of Service (QoS), good security, all-IP core network support [47, 48]. Wi-Fi and WiMAX system with the help of orthogonal frequency division multiplexing (OFDM) [5,19, 25-28], multiple access version of OFDM known as OFDMA, MIMO technology can gain speeds in a unified platform for voice communication, video communication, V2V communication. The advantage of using OFDM makes WiMAX/WiFi to overcome the multipath fading effects and still obtain a high data rate [26-28]. Benefits from the channel of MIMO system [6] can be adopted by Wi-Fi and WiMAX in order to improve the wireless channel and to double the capacity without increasing the channel bandwidth and the required power of the transmitting antenna. As a result, the spectral efficiency can also be increased at an exponential rate. MIMO system also can play an important role in vehicular communication. Massive MIMO is considered a key technology in 5G communication [49-51].

*1) Threats*

Vo-Huu et al. [52] examined the IEEE 802.11 (Wi-Fi) standard, and found that its interleaver/ convolutional scheme leaves it vulnerable to jamming attacks. Nakhila and Zhou examined how Evil Twin attacks on Wi-Fi can be detected [53]. Scarfone et al. discussed some threats that hampers the WiMAX security. The threats are radio frequency jamming, radio frequency interference, DoS, replay, man-in-the-middle, and eavesdropping attacks [54]. Kolias et al. [55] provided an extensive listing of attacks targeting the WiMAX architecture.

*2) Countermeasures*

Vo-Huu et al. proposed incorporating random, encrypted interleaving in order to protect against Wi-Fi jamming attacks [52]. Scarfone et al. [54] described technical countermeasures for WiMAX's threats such as Confidentiality and Integrity Protection, Authentication and Authorization, Client Device Security, and Patches, Upgrades, and Updates. Kolias et al. [55] also offered focusing security efforts on the first entry to the network, private key management, multicast/broadcast communications, and mesh mode traffic rerouting in order to solve the WiMAX's vulnerabilities.

### G. RFID

RFID uses radio signals to identify objects. In a RFID system, an RFID reader is able to read or identify tags attached to objects. Apart from vehicle tracking, RFID is used for toll collection, vehicle performance monitoring and vehicle routing [56-58].

*1) Threats*

Cho et al. [59] identified the two main threats to RFID communication:

  i) privacy infringement

  ii) forgery.

Cho et al. also identified potential attacks against RFID communication as Eavesdropping, Brute Force, Replay, and MITM attacks. Tyagi et al. [60] mentioned some attacks in VANET such as message spoofing, DoS attack, tracking the movement of the vehicle etc.

*2) Countermeasures*

Mohamed El Beqqal et al. [61] proposed some countermeasures in order to protect the personal data from the attacks and reduce the probability of tracking. Some protection methods mentioned in [61] are:

  a) Delegation Tree

  b) Tag Killing

  c) XOR Encryption

  d) Protocol based technique such as grouping proof protocol [62], multi-owner multi-tag transfer protocol [63], Randomized Skip Graphs-Based (RSG) RFID authentication [64] etc.

Kazuya Sakai et. al. [65] proposed a mathematical model after investigating the weak privacy in the authentication system of RFID. The proposed model in [65] can identify the probability of two tags, which are linked with respect to the number of group keys.

## III. Direct Communication

Modern day vehicles have a number of ports. These ports enable vehicles to connect to many external devices. In this way, drivers gain access to a number of services including maintenance information, on-board entertainment, and synchronization of mobile phones. In the case, an eavesdropper gets access to the vehicle ports, he/she can access the in-vehicle network and can attack the network, install virus or malware. In this section, the vulnerabilities of vehicle ports are discussed. For this, on-board diagnostics (OBD) II port, the universal serial bus (USB) port, and the vehicle charging ports are taken into consideration. Furthermore, the possible solutions to these attacks are also discussed.

### A. OBD-II Ports

OBD means on board diagnostics. OBD-II is a type of computer that is capable of monitoring the speed, mileage and other necessary data. It is connected to the check engine light. This light illuminates at the time of detecting any problem of the system. On-board diagnostics can enable digital ingress to the public data. The data may be error codes or control message. On board diagnostics can also enable access to electronic control unit (ECU) settings that can help in order to mitigate theft protection and vehicle's engine control [66, 67]. OBD generation with features are described in table 2.

| Generation | Features | Introducing Year |
|---|---|---|
| 1st | 1) Electric faults detection | 1994 |
|  | 2) Interface is manufacturer specific |  |
|  | 3) Only shows the check engine light message |  |
| 2nd | 1) Interface is universal. | 1996 |
|  | 2) Works without wire with the help of wi-fi or Bluetooth. |  |
|  | 3) High accuracy |  |

*Table 2*. OBD generation

*1) In-Vehicle Network Access Attack*

Paul Carsten discussed some threats of in-vehicle networks in a study [68]. In [68], CAN protocol was considered as crucial attack vector where CAN means controller area network. CAN protocol can behave as a broadcast network. The important feature of CAN is efficient error-checking mechanisms. The mechanism can play an important role in order to confirm that packet errors at the time of transmission can be handled proficiently. According to the Carsten et. el. [68], the threats are:

(i) Cannot able to identify CAN's inability to identify the legitimate nodes.

(ii) Packets do not carry sender address.

(iii) Packets do not carry receiver address.

(iv) No node authentication.

Miller [44] used an ECOM cable and homemade connectors for transmitting and receiving messages over CAN.

*2) Countermeasures*

There are some solutions against the vulnerabilities.

Schulze et al. [69] proposed data management system (DMS) for the purpose of monitoring communication as well as identification of malicious behavior. Ling [70] provided an algorithmic solution to DoS attack and error flag utilization of CAN protocol. According to Liu et al., OBD-II port should be separated from the in-vehicle network [23]. Oguma and his team proposed an approach using a method of attestation with encryption. This approach is able to identify the nodes existed at the startup of the system [71].

### B. USB Port

USB ports are very popular and these ports can connect phones, USB devices with entertainment files and navigation systems to the vehicle [72]. There are several types of USB ports: USB-A, USB-B, Mini USB. Micro USB, USB-C.

#### 1) Threats

USB devices persist a target for cyber threats. According to the Kaspersky Lab data, 2017, it is showed that in every 12 months around one in four users worldwide is affected by a 'local' cyber incident. These attacks are caused by USB devices [46].

#### 2) Countermeasures

According to Onishi [45] protection against USB threats is difficult because there is high amount of USB devices available in the market. Two solutions were proposed by Onishi:

a) Connect USB initially to a website for receiving a security certificate. In this case, only USB devices with certification should be allowed to be used with a vehicle.

b) Block propagation from a non-critical area to a safety critical area. In this case, even if a virus or malware gets access into a vehicle, it will have less opportunity to get into the safety critical portion of the vehicles.

## IV. Conclusion

As part of ITS, VANETs can contribute to the traffic management, road safety and drivers' comfort. In VANETs, the vehicles exchange traffic conditions, accident warnings, etc. information. Vehicle also communicate with infrastructures along the roads. Such communication facilities have introduced a number of security and privacy threats. In this review paper, the cyber security in VANET is discussed. Both remote and direct vehicle communication scenarios are taken into consideration. A number of threats such as radio frequency jamming, radio frequency interference, DoS, replay, man-in-the-middle, eavesdropping attacks, virus, malware, sniffing, fuzzing, privacy infringement, forgery are discussed. Possible safety measures are also described in this paper. In future, more security frameworks have to be designed to ensure attack-free vehicle communication systems.

## References

[1] M. Gerla, E. Lee, G. Pau and U. Lee, "Internet of vehicles: From intelligent grid to autonomous cars and vehicular clouds," *2014 IEEE World Forum on Internet of Things (WF-IoT)*, pp. 241-246, Seoul, 2014, doi: 10.1109/WF-IoT.2014.6803166.

[2] Grassi, G., Pesavento, D., Pau, G., Vuyyuru, R., Wakikawa, R. and Zhang, L., "VANET via Named Data Networking". In *Proceedings of the IEEE Conference on Computer Communications Workshops*, Castle Toronto, pp. 410-415, 27 April-2 May, 2014. doi: 10.1109/infcomw.2014.6849267.

[3] Ahmad, Farhan and Adnane, Asma, "Design of trust based context aware routing protocol in vehicular networks", *Ninth IFIP WG 11.11 International Conference on Trust Management*, 26-29 May 2015, Hamburg, Germany.

[4] M. Z. Alam, I. Adhicandra and A. Jamalipour, "Optimal Best Path Selection Algorithm for Cluster-Based Multi-Hop MIMO Cooperative Transmission for Vehicular Communications", *IEEE Transactions on Vehicular Technology*, 68 (9), pp. 8314-8321, Sept. 2019, doi: 10.1109/TVT.2019.2917695.

[5] A. K. Galib, N. Sarker and M. R. H. Mondal, "Two Compact Multiband Millimetre Wave Antennas for Wireless Communication," In *Proceedings of the 2019 IEEE International Conference on Telecommunications and Photonics (ICTP)*, Dhaka, Bangladesh, pp. 1-4, 2019. doi: 10.1109/ICTP48844.2019.9041755.

[6] Ahmad, F., Adnane, A. and N. L. Franqueira, V., "A Systematic Approach for Cyber Security in Vehicular Networks", *Journal of Computer and Communications*, 4, pp. 38-62, 2016. doi: 10.4236/jcc.2016.416004

[7] F. D. Garcia, D. Oswald, "Lock it and still lose it-on the (in) security of automotive remote keyless entry systems". In *Proceedings of the 25th USENIX Security Symposium*, 2016.

[8] Tewari A., Gupta B.B., "An Analysis of Provable Security Frameworks for RFID Security", in: *Handbook of Computer Networks and Cyber Security*, Gupta B., Perez G., Agrawal D., Gupta D. (eds). Springer, Cham, 2014.

[9] T. Glocker, T. Mantere, M. Elmusrati, "A protocol for a secure remote keyless entry system applicable in vehicles using symmetrickey cryptography". In *Proceedings of the 8th International Conference on Information and Communication Systems (ICICS), IEEE*, pp. 310–315, 2017.

[10] H.-L. Liu, J.-S. Ma, S.-Y. Zhu, Z.-J. Lu, Z.-L. Liu, "Practical contactless attacks on hitag2-based immobilizer and RKE systems". In *Proceedings of the International Conference on Computer, Communication and Network Technology*, pp. 505–512, 2018.

[11] S. Van De Beek, F. Leferink, Vulnerability of remote keyless-entry systems against pulsed electromagnetic interference and possible improvements, IEEE Transactions on Electromagnetic Compatibility, 58, pp. 1259–1265, 2016.

[12] E. Hamadaqa, A. Mars, W. Adi, S. Mulhem, "Clone-resistant vehicular RKE by deploying SUC". In *Proceedings of the 7th International Conference on Emerging Security Technologies*, pp. 221–225, 2017. doi:10.1109/ EST.2017.8090427.

[13] Q. Zhang, M. Almulla, A. Boukerche, "An improved scheme for key management of RFID in vehicular Adhoc networks", *IEEE Latin America Transactions*, 11 (6), pp. 1286– 1294, 2013.

[14] Jae Dong Lee, Hyung Jin Im, Won Min Kang, and Jong Hyuk Park, "Ubi-RKE: a rhythm key based encryption scheme for ubiquitous devices", *Mathematical Problems in Engineering*, 2014. doi:10.1155/2014/683982.

[15] Y. Li, "An Overview of the DSRC/WAVE Technology", in: *QShine 2010: Quality, Reliability, Security and Robustness in Heterogeneous Networks*, Zhang X., Qiao D. (eds), Lecture Notes of the Institute

for Computer Sciences, Social Informatics and Telecommunications Engineering, Springer, Berlin, Heidelberg", Vol. 74, pp. 544–558, 2010. doi: 10.1007/978-3-642-29222-4_38.
[16] C. Laurendeau, M. Barbeau, "Threats to security in DSRC/WAVE". In *Proceedings of the International Conference on Ad-Hoc Networks and Wireless*, pp. 266–279, 2006.
[17] S. Ucar, S. C. Ergen, O. Ozkasap, "Security vulnerabilities of IEEE 802.11p and visible light communication based platoon". In *Proceedings of the 2016 IEEE Vehicular Networking Conference,* pp. 1-4, 2016. doi:10.1109/ VNC.2016.7835972.
[18] S. Ucar, S. C. Ergen, O. Ozkasap, "IEEE 802.11p and visible light hybrid communication based secure autonomous platoon", *IEEE Transactions on Vehicular Technology*, 67, pp.8667–8681, 2018.
[19] M. R. H. Mondal, "Impact of Spatial Sampling Frequency Offset and Motion Blur on Optical Wireless Systems using Spatial OFDM", EURASIP Journal on Wireless Communications and Networking, Springer, DOI: 10.1186/s13638-016-0741-y, Oct 2016.
[20] S. Biswas, J. Mišić, V. Mišić, "DDoS attack on WAVE-enabled VANET through synchronization". In Proceedings of the *Globecom 2012: Communication and Information System Security Symposium*, pp. 1079–1084, 2012.
[21] W. Whyte, J. Petit, V. Kumar, J. Moring, R. Roy, "Threat and countermeasures analysis for WAVE service advertisement". In *Proceedings of the IEEE 18th International Conference on Intelligent Transportation Systems*, pp. 1061–1068, 2015.
[22] N. Lyamin, A. Vinel, M. Jonsson, J. Loo, "Real-time detection of denial-of-service attacks in IEEE 802.11p vehicular networks", *IEEE Communications Letters*, Vol. 18, pp.110– 113, 2014.
[23] H. Nguyen-Minh, A. Benslimane, A. Rachedi, "Jamming detection on 802.11p under multi-channel operation in vehicular networks". In *Proceedings of the 2015 IEEE 11th International Conference on Wireless and Mobile Computing, Networking and Communications,* pp. 764–770, 2015. doi:10.1109/WiMOB.2015.7348039.
[24] N. Ghambir, P. Sharma, "A hybrid approach for intelligent communication and performance analysis over DSRC VANET". In *Proceedings of the IEEE International Conference on Information, Communication, Instrumentation and Control*, 2017. doi:10.1109/ISS1.2017.8389327.
[25] M. R. H. Mondal and Jean Armstrong, "Analysis of the effect of vignetting on MIMO optical wireless systems using spatial OFDM", *Journal of lightwave technology*, 32 (5), pp. 922-929, 2013.
[26] M. R. H. Mondal and Satya P. Majumder, "Analytical Performance Evaluation of Space Time Coded MIMO OFDM Systems Impaired by Fading and Timing Jitter", Journal of Communications (JCM), ISSN: 1796-2021, vol. 4, issue. 6, July 2009.
[27] M. M. H. Mishu, M. R. H. Mondal, "Effectiveness of Filter Bank Multicarrier Modulation for 5G Wireless Communications", International Conference on Advances in Electronics Engineering (ICAEE), Dhaka, Bangladesh, Sep. 2017.
[28] N. Sarker, M. A. Islam, and M. R. H. Mondal, "Two Novel Multiband Centimetre-Wave Patch Antennas for a Novel OFDM Based RFID System", Journal of Communications (JCM), ISSN: 1796-2021, vol. 13, no. 6, Jun. 2018.
[29] Abboud, K., Omar, H. A., & Zhuang, W., "Interworking of DSRC and Cellular Network Technologies for V2X Communications: A Survey", *IEEE Transactions on Vehicular Technology*, 65(12), pp. 9457–9470, 2016. doi:10.1109/tvt.2016.2591558
[30] J. Cao, M. Ma, H. Li, Y. Zhang, Z. Luo, "A survey on security aspects for LTE and LTE-A networks", *IEEE Communications Surveys and Tutorials,* Vol. 16, pp. 283–302, 2014.
[31] K. Kaur, A. S. Sharma, H. S. Sohal and A. Kaur, "Adaptive Random Key Scheme for Authentication and Key Agreement (ARKS-AKA) for efficient LTE security". In *Proceedings of the 2015 2nd International Conference on Recent Advances in Engineering & Computational Sciences (RAECS)*, Chandigarh, pp. 1-6, 2015. doi: 10.1109/RAECS.2015.7453422.
[32] Y. Park and T. Park, "A Survey of Security Threats on 4G Networks". In Proceedings of the IEEE Globecom Workshops, pp.1-6, November 2007.
[33] Borgaonkar, R., Hirschi, L., Park, S., & Shaik, A., "New Privacy Threat on 3G, 4G, and Upcoming 5G AKA Protocols", In *Proceedings on Privacy Enhancing Technologies*, 2019(3), pp. 108-127, 2019. doi:10.2478/popets-2019-0039
[34] 3rd Generation Partnership Project; Technical Specification Group Services and System Aspects; Rationale and track of security decisions in Long Term Evolved (LTE) RAN/3GPP System Architecture Evolution (SAE), (Rel 9), 3GPP TR 33.821 V9.0.0 June 2009.
[35] D. Rupprecht, K. Jansen, C. Pöpper, "Putting LTE security functions to the test: a framework to evaluate implementation correctness". In *Proceedings of the 10th USENIX Conference on Offensive Technologies,* pp. 40–51, 2016.
[36] M. Liyanage, I. Ahmad, M. Ylianttila, A. Gurtov, A. B. Abro and E. M. de Oca, "Leveraging LTE security with SDN and NFV". In *Proceedings of the 2015 IEEE 10th International Conference on Industrial and Information Systems (ICIIS)*, Peradeniya, pp. 220-225, 2015. doi: 10.1109/ICIINFS.2015.7399014.
[37] J. Cichonski, J. M. Franklin, M. Bartock, "Guide to LTE Security, Technical Report", NIST, 2017. doi: 10.6028/NIST.SP.800-187.
[38] Manpreet and J. Malhotra, "ZigBee technology: Current status and future scope". In *Proceedings of the 2015 International Conference on Computer and Computational Sciences (ICCCS),* Noida, pp. 163-169, 2015. doi: 10.1109/ICCACS.2015.7361343.
[39] Yu Lei and Jian Wu, "Study of applying ZigBee technology into forward collision warning system (FCWS) under low-speed circumstance". In *Proceedings of the 2016 25th Wireless and Optical Communication Conference (WOCC)*, Chengdu, pp. 1-4, 2016. doi: 10.1109/WOCC.2016.7506630.
[40] S. Pawade, S. Shah, D. Tijare, "Zigbee based intelligent driver assistance system", *International Journal of*

*Engineering Research and Applications*, Vol. 3, pp.1463–1468, 2013.

[41] N. Vidgren, K. Haataja, J. L. Patiño-Andres, J. J. Ramírez-Sanchis and P. Toivanen, "Security Threats in ZigBee-Enabled Systems: Vulnerability Evaluation, Practical Experiments, Countermeasures, and Lessons Learned". In *Proceedings of the 46th Hawaii International Conference on System Sciences*, Wailea, Maui, HI, pp. 5132-5138, 2013. doi: 10.1109/HICSS.2013.475.

[42] Olawumi, Olayemi & Haataja, Keijo & Asikainen, M. & Vidgren, Niko & Toivanen, Pekka, "Three Practical Attacks Against ZigBee Security: Attack Scenario Definitions, Practical Experiments, Countermeasures, and Lessons Learned". In *Proceedings of the 14th International Conference on Hybrid Intelligent Systems*, HIS 2014. doi:10.1109/HIS.2014.7086198.

[43] J. Cache, J. Wright, and V. Liu, Hacking Exposed Wireless: Wireless Security Secrets and Solutions, McGraw-Hill, Second Edition, Jul. 2010.

[44] C. Valasek and C. Miller, "Adventures in automotive networks and control units", *DEFCON* 23, 2015.

[45] Hirofumi Onishi, Kelly Wu, Kazuo Yoshida, Takeshi Kato, "Approaches for vehicle cyber-security in the US", *International Journal of Automotive Engineering*, 8 (2017), pp. 1-6, 2017.

[46] M. Cheah, S. A. Shaikh, O. Haas, A. Ruddle, "Towards a systematic security evaluation of the automotive Bluetooth interface", *Vehicular Communications*, Vol. 9, pp. 8–18, 2017.

[47] Pranav Kumar Singh, Sunit Kumar Nandi, Sukumar Nandi, "A tutorial survey on vehicular communication state of the art, and future research directions", *Vehicular Communications*, Vol.18, 100164, 2019. doi: 10.1016/j.vehcom.2019.100164.

[48] K. Tanuja, T. M. Sushma, M. Bharathi, K. H. Arun, "A survey on VANET technologies", *International Journal of Computer Applications*, 121 (18), July 2015.

[49] Bharati, S., Podder, P. "Adaptive PAPR Reduction Scheme for OFDM Using SLM with the Fusion of Proposed Clipping and Filtering Technique in Order to Diminish PAPR and Signal Distortion", *Wireless Personal Communi*cations, Springer Publisher, 2020. https://doi.org/10.1007/s11277-020-07323-0.

[50] Subrato Bharati, Prajoy Podder, Niketa Gandhi, Ajith Abraham, "Realization of MIMO Channel Model for Spatial Diversity with Capacity and SNR Multiplexing Gains", *International Journal of Computer Information Systems and Industrial Management Applications*, Vol.12, pp. 66-81, 2020.

[51] Prajoy Podder, Subrato Bharati, Md. Robiul Alam Robel, Md. Raihan Al- Masud and Mohammad Atikur Rahman, "Uplink and Downlink Spectral Efficiency Estimation for Multi Antenna MIMO User", in: *Innovations in Bio-Inspired Computing and Applications*, Abraham, A., Panda, M., Pradhan, S., Garcia-Hernandez, L., Ma, K. (eds.). Advances in Intelligent Systems and Computing, Vol. 1180, Springer, Cham, 2020.

[52] Triet D. Vo-Huu, Tien D. Vo-Huu, Guevara Noubir, Interleaving jamming in wi-fi networks. In *Proceedings of the 9th ACM Conference on Security & Privacy in Wireless and Mobile Networks*, 2016, pp. 31–42. doi:10.1145/2939918.2939935.

[53] O. Nakhila, E. Dondyk, M. F. Amjad, C. Zou, User-side wi-fi evil twin attack detection using random wireless channel monitoring. *In Proceedings of the 12th Annual IEEE Consumer Communications and Networking Conference*,2015.doi:10.1109/MILCOM.2016.7795501

[54] K. Scarfone, C. Tibbs, M. Sexton, Guide to securing wimax wireless communications: recommendations of the national institute of standards and technology, Technical Report, NIST, 2010. doi:10.6028/NIST.SP.800-127.

[55] C. Kolias, G. Kambourakis, S. Gritzalis, "Attacks and countermeasures on 802.16: analysis and assessment", *IEEE Communications Surveys & Tutorials*, 15, pp. 487–514, 2013.

[56] F. Moradi, H. Mala, B. T. Ladani, "Security analysis and strengthening of an RFID lightweight authentication protocol suitable for VANETs", *Wireless Personal Communications*, 83 (2015), pp. 2607–2621.

[57] Y. P. Liao, C. M. Hsiao, "A secure ECC-based RFID authentication scheme integrated with ID-verifier transfer protocol", *Ad Hoc Networks*, 18 (2014), pp.133–146, 2014.

[58] V. R. Vijaykumar, S. Elango, "Hardware implementation of tag reader mutual authentication protocol for RFID systems", *Integration, the VLSI Journal,* 47, pp.123–129, 2014.

[59] J. S. Cho, Y. S. Jeong, S. O. Park, "Consideration on the brute-force attack cost and retrieval cost: a hash-based radio-frequency identification (RFID) tag mutual authentication protocol", *Computers & Mathematics with Applications*, 69(1), pp. 58–65, January 2015.

[60] Amit Kumar Tyagi, Meghna Manoj Nair, Sreenath Niladhuri and Ajith Abraham, "Comparative Analysis of Multi-round Cryptographic Primitives based Lightweight Authentication Protocols for RFID-Sensor Integrated MANETs", *Journal of Information Assurance and Security (ISSN 1554-1010)*, Volume 15, pp. 001-016, 2020.

[61] Mohamed El Beqqal, Mostafa Azizi, "Review on security issues in RFID systems", *Advances in Science, Technology and Engineering Systems Journal*, 2(6), pp. 194-202, 2017. doi: 10.25046/aj020624.

[62] Sundaresan, S, Doss, R. ; Piramuthu, S. ; Wanlei Zhou, "A Robust Grouping Proof Protocol for RFID EPC C1G2 Tags", *IEEE Transactions on Information Forensics and Security*, 9(6), 2014. https://doi.org/10.1109/TIFS.2014.231633

[63] Sundaresan, S. Doss, R., Wanlei Zhou, "Secure ownership transfer in multi-tag/multi-owner passive RFID systems". In *Proceedings of the IEEE Global Communications Conference (GLOBECOM)*, 2013. doi: 10.1109/GLOCOM.2013.683151

[64] Yudai Komori, Kazuya Sakai, Satoshi Fukumoto, "Fast and secure tag authentication in large-scale RFID systems using skip graphs", *Computer Communications*, Vol. 116, pp. 177-189, 2018.

[65] Kazuya Sakai, Min-Te Sun, Wei-Shinn Ku, Ten H Lai, "On The Performance Bound of Structured Key-Based RFID Authentication". In *Proceedings of the IEEE International Conference on Pervasive Computing and Communications (PerCom)*, pp. 1-10, 2019.

[66] A. Yadav, G. Bose, R. Bhange, K. Kapoor, N. Iyengar and R. Caytiles, "Security, Vulnerability and Protection of Vehicular On-board Diagnostics", *International Journal of Security and Its Applications,* 10 (4), pp.405-422, 2016. doi: 10.14257/ijsia.2016.10.4.36.

[67] https://www.escrypt.com/fileadmin/escrypt/pdf/Whitepaper/OBD_Open_Barn_Door_Security_Vulnerabilities_and_Protections_for_Vehicular_On_Board_Diagnosis.pdf (Last Accessed on 28 May 2020).

[68] P. Carsten, M. Yampolskiy, T. R. Andel, J. T. Mcdonald, Invehicle networks: attacks, vulnerabilities, and proposed solutions. In *Proceedings of the 10th Annual Cyber and Information Security Research Conference*, 2015.

[69] S. Schulze, M. Pukall, G. Saake, T. Hoppe, and J. Dittmann, "On the need of automotive data management in automotive systems". In *Proceedings of the GI-Fachtagung Datenbanksysteme für Business, Technologie und Web (BTW '09)*, Lecture Notes in Informatics, pp. 217–227, Gesellschaft für Informatik (GI), March 2009.

[70] Ling, Congli, and Dongqin Feng. "An Algorithm for Detection of Malicious Messages on CAN Buses". In *Proceedings of the 2012 National Conference on Information Technology and Computer Science*, Atlantis Press, 2012.

[71] H. Oguma, A. Yoshioka, M. Nishikawa, R. Shigetomi, A. Otsuka and H. Imai, "New Attestation Based Security Architecture for In-Vehicle Communication". In *Proceedings of the IEEE GLOBECOM 2008 - 2008 IEEE Global Telecommunications Conference*, New Orleans, LO, pp. 1-6, 2008.

[72] F. He, "USB Port and power delivery: An overview of USB port interoperabiliy". In *Proceedings of the 2015 IEEE Symposium on Product Compliance Engineering (ISPCE)*, Chicago, IL, pp. 1-5, 2015. doi: 10.1109/ISPCE.2015.7138710.

[73] Usman Ali Khan and Sang Sun Lee, "Multi-Layer Problems and Solutions in VANETs: A Review", *Electronics*, MDPI Publisher, 8(2), 204, 2019. doi: 10.3390/electronics8020204.

[74] Muhammad Arif, Guojun Wang, Md Zakirul Alam Bhuiyan, Tian Wang, Jianer Chen, "A survey on security attacks in VANETs: Communication, applications and challenges", *Vehicular Communications*, Volume 19, 100179, October 2019.

## Author Biographies

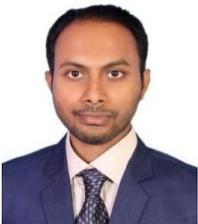

**Subrato Bharati** received his B.S degree in Electrical and Electronic Engineering from Ranada Prasad Shaha University, Narayanganj-1400, Bangladesh. He is currently working as a research assistant at the Institute of Information and Communication Technology, Bangladesh University of Engineering and Technology, Dhaka, Bangladesh. He is a regular reviewer of ISA Transactions, Elsevier; Array, Elsevier; Vehicular Communications, Elsevier; Journal of Systems Architecture, Elsevier; Cognitive Systems Research, Elsevier; Soft Computing, Springer; Data in Brief, Elsevier, Wireless Personal Communications, Springer; Informatics in Medicine Unlocked, Elsevier. He was a member of the Scientific Committee of International Conference on Cyber Security and Computer Science (ICONCS 2020). He was also a reviewer of some international conferences. He is the guest editor of Special Issue on Development of Advanced Wireless Communications, Networks and Sensors in American Journal of Networks and Communications. His research interest includes bioinformatics, medical image processing, pattern recognition, deep learning, wireless communications, data analytics, machine learning, neural networks, distributed sensor networks, parallel and distributed computing computer networking, digital signal processing, telecommunication and feature selection. He published several IEEE, Springer reputed conference papers and also published several journals paper, Springer and Elsevier Book chapters.

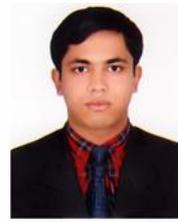

**Prajoy Podder** worked as a Lecturer in the Department of Electrical and Electronic Engineering in Ranada Prasad Shaha University, Narayanganj-1400, Bangladesh. He completed B.Sc. (Engg.) degree in Electronics and Communication Engineering from Khulna University of Engineering & Technology, Khulna-9203, Bangladesh. He is currently pursuing M.Sc. (Engg.) degree in Institute of Information and Communication Technology from Bangladesh University of Engineering and Technology, Dhaka-1000, Bangladesh. He is a researcher in the Institute of Information and Communication Technology, Bangladesh University of Engineering & Technology, Dhaka-1000, Bangladesh. He is regular reviewer of Data in Brief, Elsevier and Frontiers of Information Technology and Electronic Engineering, Springer, ARRAY, Elsevier. He is the lead guest editor of Special Issue on Development of Advanced Wireless Communications, Networks and Sensors in American Journal of Networks and Communications. His research interest includes machine learning, pattern recognition, neural networks, computer networking, distributed sensor networks, parallel and distributed computing, VLSI system design, image processing, embedded system design, data analytics. He published several IEEE conference papers, journals, Springer and Elsevier Book Chapters.

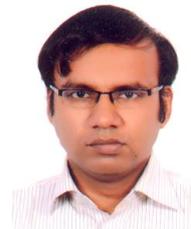

**M. Rubaiyat Hossain Mondal**, PhD is currently working as an Associate Professor in the Institute of Information and Communication Technology (IICT) at Bangladesh University of Engineering and Technology (BUET), Bangladesh. He received his Bachelor's degree and Master's degree in Electrical and Electronic Engineering from BUET. He joined IICT, BUET as a faculty member in 2005. From 2010 to 2014 he was with the Department of Electrical and Computer Systems Engineering (ECSE) of Monash University, Australia from where he obtained his PhD in 2014. He has authored a number of articles in reputed journals published by IEEE, Elsevier, De Gruyter, IET, Springer, PLOS, Wiley and MDPI publishers. He has also authored a number of IEEE conference papers including GLOBECOM 2010 in USA, and presented papers in IEEE conferences in Australia, South Korea, and Bangladesh. In addition, he has coauthored a number of book chapters of reputed publishers which are now in press. He is an active reviewer of several journals published by IEEE, Elsevier and Springer. He was a member of the Technical Committee of different IEEE R10 International conferences. His research interest includes artificial intelligence, bioinformatics, image processing, wireless communication and optical wireless communication.

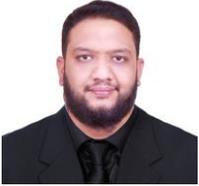

**Md. Robiul Alam Robel** received his Bachelor degree in Computer Science and Engineering from Comilla University, Comilla, Bangladesh. He published several papers in springer scopus indexed conference. His research interest includes machine learning, soft computing, cloud computing, natural language processing etc.